# Is the electron magnetic moment unique?

V. A. Golovko

Moscow Polytechnic University

Bolshaya Semenovskaya 38, Moscow 107023, Russia

E-mail: fizika.mgvmi@mail.ru

## Abstract

There exist two methods for finding the magnetic moment of the electron. The first method employed in quantum electrodynamics consists in calculating the energy of the electron placed in a constant magnetic field, the extra energy due to the field being proportional to the magnetic moment. It is also possible to use the second method proceeding from the fact that the asymptotic form of the vector potential at infinity is proportional to the magnetic moment. If the electron were point-like, both the methods would yield identical results. In the present paper is studied the magnetic field created by the electron in hydrogen-like ions, which enables one to find the electron magnetic moment by the second method. The electron magnetic moment in this case proves to be different in different states of the electron in the Coulomb field of the ions and, moreover, is distinct from the magnetic moment calculated by the first method. The results of the paper show that the electron is not small and is deformable under action of external fields.

# 1. Introduction

The magnetic moment of the electron with its anomalous part is calculated in quantum electrodynamics (QED). The electron magnetic moment μ with the Schwinger correction is

$$\mu = \frac{e\hbar}{2m_e c}\left(1 + \frac{\alpha}{2\pi}\right), \tag{1.1}$$

where $\alpha = e^2/\hbar c$ is the fine-structure constant. It should be stressed that the magnetic moment in QED is obtained by considering the electron placed in a constant magnetic field, in which case the extra energy of the electron will be proportional to the magnetic moment.

There is another way to find out the magnetic moment of a system. The asymptotic form of the vector potential **A** as $r \to \infty$ is proportional to the magnetic moment **μ** of the system. If **μ** is directed along the z-axis, there will be only one nonzero component of the asymptotic vector **A**, namely, the φ-component $A_\varphi$ in spherical coordinates and [1]

$$A_\varphi = \frac{\mu}{r^2}\sin\vartheta. \tag{1.2}$$

It should be remarked that Eq. (1.2) holds in the gauge in which div **A** = 0 by virtue of Eq. (43.3) of [1].

If the electron were point-like, both the methods would yield identical results. A point-like object having a mass, charge and spin cannot be a physical object; the point-like object is a mathematical abstraction pure and simple (we shall return to this question in the concluding section). If the electron extends in space, it cannot be absolutely rigid because an absolutely rigid body is an abstraction as well. Once the electron is deformable, its structure and thereby its magnetic moment should change under the action of external forces, which amounts to saying that the electron magnetic moment is not unique but depends upon external conditions.

In the present paper we shall consider the electron in hydrogen or hydrogen-like ions and demonstrate that its magnetic moment is different in different states. The results of the paper show that the electron has finite dimensions and is deformable. Moreover the results suggest that the sizes of the electron are rather noticeable.

# 2. The vector potential

In a stationary state the vector potential **A** satisfies the well-known equation [1]

$$\nabla^2 \mathbf{A} = -\frac{4\pi e}{c}\mathbf{j}, \tag{2.1}$$

where **j** is the current density (we do not include the charge $e$ into this definition). We also assume that div **A** = 0, which holds for all examples considered in the present paper (see below). Consequently, Eq. (1.2) can be used since div **A** = 0. In the case of the Dirac equation one has

$$\mathbf{j} = \psi^* \boldsymbol{\alpha} \psi \,, \tag{2.2}$$

where $\boldsymbol{\alpha}$ are the standard matrices [2].

The bispinor $\psi$ is of the form

$$\psi = \begin{pmatrix} \psi_1 \\ \psi_2 \\ \psi_3 \\ \psi_4 \end{pmatrix}. \tag{2.3}$$

We have now from (2.2) in the standard representation that

$$j_x = \psi_1^*\psi_4 + \psi_2^*\psi_3 + \psi_3^*\psi_2 + \psi_4^*\psi_1, \quad j_y = -i\psi_1^*\psi_4 + i\psi_2^*\psi_3 - i\psi_3^*\psi_2 + i\psi_4^*\psi_1,$$

$$j_z = \psi_1^*\psi_3 - \psi_2^*\psi_4 + \psi_3^*\psi_1 - \psi_4^*\psi_2 \,. \tag{2.4}$$

These components can be written in spherical coordinates according to

$$j_r = j_x\sin\vartheta\cos\varphi + j_y\sin\vartheta\sin\varphi + j_z\cos\vartheta, \quad j_\vartheta = j_x\cos\vartheta\cos\varphi + j_y\cos\vartheta\sin\varphi - j_z\sin\vartheta,$$

$$j_\varphi = -j_x\sin\varphi + j_y\cos\varphi \,. \tag{2.5}$$

In the present paper we shall study the electron in the Coulomb field of hydrogen-like ions with charge $-Ze$. In this case the bispinor $\psi$ with quantum numbers $j$, $l$ and $m$ is [2]

$$\psi = \begin{pmatrix} f(r)\Omega_{jlm} \\ (-1)^{\frac{1+l-l'}{2}} g(r)\Omega_{jl'm} \end{pmatrix}, \tag{2.6}$$

where $l = j \pm \frac{1}{2}$, $l' = 2j - l$ and the spherical harmonic spinors $\Omega_{jlm}$ are

$$\Omega_{l+1/2,l,m} = \begin{pmatrix} \sqrt{\dfrac{j+m}{2j}}Y_{l,m-1/2} \\ \sqrt{\dfrac{j-m}{2j}}Y_{l,m+1/2} \end{pmatrix}, \quad \Omega_{l-1/2,l,m} = \begin{pmatrix} -\sqrt{\dfrac{j-m+1}{2j+2}}Y_{l,m-1/2} \\ \sqrt{\dfrac{j+m+1}{2j+2}}Y_{l,m+1/2} \end{pmatrix} \tag{2.7}$$

for the two possible values $j = l \pm \frac{1}{2}$ for a given $l$. We define the spherical harmonic functions $Y_{lm}$ in (2.7) according to [3]. Henceforward, the notation of [2] is implied; in particular, we set $\hbar = c = 1$. The notation used will be explained in the next section.

When Eqs. (2.6) and (2.7) are placed in (2.4), one obtains that $j_z = 0$. Subsequent substitution into (2.5) yields that $j_r = j_\theta = 0$ (this is obvious in advance for the symmetry in question) and

$$j_\varphi = P - Q, \tag{2.8}$$

where

$$P = (-1)^{l+1} i e^{-2im\varphi} f(r) g(r) \sqrt{\frac{(j+m)(j+m+1)}{j(j+1)}} \, Y_{l,m\mp1/2} Y_{l\pm1,m\pm1/2} \,,$$

$$Q = (-1)^{l} i e^{-2im\varphi} f(r) g(r) \sqrt{\frac{(j-m)(j-m+1)}{j(j+1)}} \, Y_{l,m\pm1/2} Y_{l\pm1,m\mp1/2} \,, \qquad (2.9)$$

in which the signs correspond to the signs in $j = l \pm \frac{1}{2}$. The noteworthy fact is that the quantities $P$ and $Q$ are real and do not contain the angle $\varphi$.

Seeing that the component $j_\varphi$ alone is different from zero, Eq.(2.1) in spherical coordinates acquires the form

$$\left( \nabla^2 \mathbf{A} \right)_\varphi = -4\pi e j_\varphi \,, \qquad (2.10)$$

or

$$\frac{1}{r^2} \frac{\partial}{\partial r} \left( r^2 \frac{\partial A_\varphi}{\partial r} \right) + \frac{1}{r^2} \frac{\partial}{\partial \vartheta} \left[ \frac{1}{\sin\vartheta} \frac{\partial}{\partial \vartheta} \left( \sin\vartheta A_\varphi \right) \right] = -4\pi e j_\varphi \,, \qquad (2.11)$$

because $j_\varphi$ is independent of $\varphi$. Since $A_\varphi$ is also independent of $\varphi$, one has div $\mathbf{A} = 0$.

### 3. The magnetic moment of the electron

According to the Dirac equation, states of the electron in a Coulomb-like field are described in the following way. The quantum number $n$ is analogous to the principal quantum number of the non-relativistic theory where it determines the energy of the electron. The angular momentum operator $\mathbf{J}$ is the sum of the orbital angular momentum operator $\mathbf{L} = [\mathbf{r}\cdot\mathbf{p}]$ with the momentum $\mathbf{p} = -i\nabla$ and the spin momentum operator $\frac{1}{2}\boldsymbol{\sigma}$:

$$\mathbf{J} = \mathbf{L} + \tfrac{1}{2}\boldsymbol{\sigma}. \qquad (3.1)$$

The square of the angular momentum is $\mathbf{J}^2 = j(j + 1)$ where $j$ is the total angular momentum quantum number, and the quantum number $m$ determines the projection of the vector $\mathbf{J}$ on the z-axis ($m = -j, -j + 1 \ldots j - 1, j$). It is seen from (3.1) that $j = l \pm \frac{1}{2}$ while the orbital quantum number $l$ follows from the relation $\mathbf{L}^2 = l(l + 1)$ ($l = 0, 1, 2, n - 1$). The quantum number $\kappa$ points to one of two possible types of solutions to the equations, namely,

$$\kappa = \begin{cases} -j(j + \tfrac{1}{2}) = -(l+1) & \text{for } j = l + \tfrac{1}{2}, \\ +j(j + \tfrac{1}{2}) = l & \text{for } j = l - \tfrac{1}{2}. \end{cases} \qquad (3.2)$$

The radial quantum number $n_r = n - |\kappa|$ determines energy levels ($n = 1, 2, \ldots$) in view of Eq. (36.10) of [2].

The functions $f(r)$ and $g(r)$ that figure in Eq. (2.6) are given in Eq. (36.11) of [2]. Here we do not write down the last equation in its general form but we shall apply it separately for different stationary states.

First, we consider the ground state $1s_{1/2}$ for which $n = 1$, $l = 0$, $j = m = \frac{1}{2}$, $n_r = 0$, $\kappa = -1$. In this state the orbital rotation is absent and the magnetic field is completely due to the magnetic moment of the electron. If $n_r = 0$, the functions $f(r)$ and $g(r)$ are simple in form:

$$f(r) = \sqrt{m_e + \varepsilon} A r^{\gamma-1} e^{-\lambda r}, \quad g(r) = -\sqrt{m_e - \varepsilon} A r^{\gamma-1} e^{-\lambda r}. \tag{3.3}$$

Hereinafter

$$\gamma = \sqrt{\kappa^2 - (Z\alpha)^2}, \ \lambda = \sqrt{m_e^2 - \varepsilon^2}, \tag{3.4}$$

where $\varepsilon$ is the energy of the state. The normalization factor $A$ can be readily found from the normalization condition

$$\int_0^\infty \left( f^2 + g^2 \right) r^2 dr = 1, \tag{3.5}$$

which yields

$$A^2 = \frac{(2\lambda)^{2\gamma+1}}{2m_e \Gamma(2\gamma+1)}. \tag{3.6}$$

It is helpful to remark that Eqs. (3.3) and (3.6) are valid for any state where $n_r = 0$.

It follows from (2.8) – (2.9) for the present example that $Q = 0$ and

$$j_\varphi = P = -ie^{-i\varphi} fg \sqrt{\frac{8}{3}} Y_{00} Y_{11} = -\frac{1}{2\pi} fg \sin\vartheta. \tag{3.7}$$

Substituting this into (2.11) gives

$$\frac{1}{r^2} \frac{\partial}{\partial r} \left( r^2 \frac{\partial A_\varphi}{\partial r} \right) + \frac{1}{r^2} \frac{\partial}{\partial \vartheta} \left[ \frac{1}{\sin\vartheta} \frac{\partial}{\partial \vartheta} \left( \sin\vartheta A_\varphi \right) \right] = 2efg \sin\vartheta. \tag{3.8}$$

We seek a solution to the equation in the form

$$A_\varphi = e\zeta(r)\sin\vartheta, \tag{3.9}$$

and obtain the following equation for $\zeta(r)$

$$\frac{d}{dr} \left( r^2 \frac{d\zeta}{dr} \right) - 2\zeta = 2r^2 fg. \tag{3.10}$$

Two solutions of the complementary homogeneous equation are $r$ and $1/r^2$, so that the solution of Eq. (3.10) can be sought by variation of constants:

$$\zeta(r) = C_1(r)r + \frac{C_2(r)}{r^2}. \tag{3.11}$$

If one introduces (3.11) into (3.10), one will obtain

$$\frac{dC_1}{dr} = \frac{2}{3} fg, \quad \frac{dC_2}{dr} = -\frac{2}{3} r^3 fg .\tag{3.12}$$

We integrate the first equation of (3.12) upon making use of (3.3):

$$C_1 = -\frac{2}{3} \lambda A^2 I_{2\gamma-2} + \overline{C_1},\tag{3.13}$$

where $\overline{C_1}$ is a constant and the value of $\lambda$ of (3.4) is taken into account.

From here on we introduce the integrals

$$I_p(r) = \int_\infty^r x^p e^{-2\lambda x} dx = \frac{-1}{(2\lambda)^{p+1}} \Gamma(p+1, 2\lambda r),\tag{3.14}$$

where $\Gamma(a,x)$ is the incomplete gamma function. These integrals satisfy the recurrence formula

$$I_p = \frac{2\lambda}{p+1} I_{p+1} + \frac{r^{p+1}}{p+1} e^{-2\lambda r} .\tag{3.15}$$

If $r \to 0$, one has

$$I_p = -\frac{\Gamma(p+1)}{(2\lambda)^{p+1}} + \frac{r^{p+1}}{p+1} + O(r^{p+2}),\tag{3.16}$$

and if $r \to \infty$, one has

$$I_p = -\frac{r^p}{2\lambda} e^{-2\lambda r} \left[ 1 + O\left(\frac{1}{r}\right) \right].\tag{3.17}$$

We revert to Eq. (3.13). We must put $\overline{C_1} = 0$; otherwise $\zeta(r)$ of (3.11) will increase with $r$. Now the first part of $\zeta(r)$ falls off exponentially as $r \to \infty$ according to (3.17).

The second equation of (3.12) can be treated analogously and Eq. (3.11) leads to

$$\zeta(r) = \frac{2}{3} \lambda A^2 \left( -rI_{2\gamma-2} + \frac{1}{r^2} I_{2\gamma+1} \right) + \frac{\overline{C_2}}{r^2} .\tag{3.18}$$

The constant $\overline{C_2}$ is to be found from the requirement that $\zeta(r)$ be finite as $r \to 0$. In view of (3.16) this gives

$$\overline{C_2} = \frac{2}{3} \lambda A^2 \frac{\Gamma(2\gamma+2)}{(2\lambda)^{2\gamma+2}} = \frac{2\gamma+1}{6m_e},\tag{3.19}$$

where we have used (3.6) and the formula $\Gamma(a+1) = a\Gamma(a)$.

It follows now from (3.18) and (3.9) that

$$A_\varphi \to \frac{e(2\gamma+1)}{6m_e r^2} \sin\vartheta ,\tag{3.20}$$

as $r \to \infty$. Upon comparing this with (1.2) we obtain the magnetic moment of the electron in the state $1s_{1/2}$:

$$\mu = \frac{e\hbar}{6m_e c}(2\gamma + 1) \approx \frac{e\hbar}{2m_e c}\left(1 - \frac{(Z\alpha)^2}{3}\right), \qquad (3.21)$$

where $\hbar$ and $c$ are written down explicitly by analogy with (1.1). The last formula is got by expanding $\gamma$ of (3.4) in powers of $\alpha$ and retaining the first two terms.

We see that the magnetic moment of the electron in the state $1s_{1/2}$ in an atom has an anomalous part but this part is distinct from the one of the same electron placed in a constant magnetic field, the latter being given by Eq. (1.1).

Next, we turn to the state $2s_{1/2}$ where $n = 2$, $l = 0$, $j = m = \frac{1}{2}$, $n_r = 1$, $\kappa = -1$. The functions $f(r)$ and $g(r)$ for this state can be taken from Eq. (36.11) of [2]:

$$f(r) = \sqrt{m_e + \varepsilon}\, A r^{\gamma - 1} e^{-\lambda r}\left[\frac{Z\alpha m_e}{\lambda} - \left(\frac{Z\alpha m_e}{\lambda} + 1\right)\frac{2\lambda r}{2\gamma + 1}\right],$$

$$g(r) = -\sqrt{m_e - \varepsilon}\, A r^{\gamma - 1} e^{-\lambda r}\left[\frac{Z\alpha m_e}{\lambda} + 2 - \left(\frac{Z\alpha m_e}{\lambda} + 1\right)\frac{2\lambda r}{2\gamma + 1}\right] \qquad (3.22)$$

with

$$A^2 = \frac{(2\gamma + 1)(2\lambda)^{2\gamma + 2}}{8Z\alpha m_e^2\left(\dfrac{Z\alpha m_e}{\lambda} + 1\right)\Gamma(2\gamma + 1)} \cdot \qquad (3.23)$$

It follows from (2.8) the same result as in (3.7) and consequently Eqs. (3.9)–(3.12) remain valid for the state under consideration. Integrating the first equation of (3.12) upon making use of (3.22) and (3.4) yields now that

$$C_1 = -\frac{2\lambda}{3} A^2\left[\frac{Z\alpha m_e}{\lambda}\left(\frac{Z\alpha m_e}{\lambda} + 2\right)I_{2\gamma - 2} - \left(\frac{Z\alpha m_e}{\lambda} + 1\right)^2\frac{4\lambda}{2\gamma + 1}I_{2\gamma - 1} + \left(\frac{Z\alpha m_e}{\lambda} + 1\right)^2\frac{4\lambda^2}{(2\gamma + 1)^2}I_{2\gamma}\right],$$

$$(3.24)$$

where the constant of integration is put equal to zero as above. The second equation of (3.12) can be treated analogously and Eq. (3.11) leads to

$$\zeta(r) = C_1(r) r$$

$$+ \frac{2\lambda A^2}{3r^2}\left[\frac{Z\alpha m_e}{\lambda}\left(\frac{Z\alpha m_e}{\lambda} + 2\right)I_{2\gamma + 1} - \left(\frac{Z\alpha m_e}{\lambda} + 1\right)^2\frac{4\lambda}{2\gamma + 1}I_{2\gamma + 2} + \left(\frac{Z\alpha m_e}{\lambda} + 1\right)^2\frac{4\lambda^2}{(2\gamma + 1)^2}I_{2\gamma + 3}\right] + \frac{\overline{C}_2}{r^2}$$

$$(3.25)$$

The constant $\overline{C}_2$ is to be found from the requirement that $\zeta(r)$ be finite as $r \to 0$. In view of (3.16) and (3.23) this gives

$$\widetilde{\mu} = \frac{\lambda(2\gamma+1)}{6Z\alpha m_e\left(\dfrac{Z\alpha m_e}{\lambda}+1\right)}\left[\frac{Z\alpha m_e}{2\lambda}\left(\frac{Z\alpha m_e}{\lambda}+2\right)(2\gamma+1)-\left(\frac{Z\alpha m_e}{\lambda}+1\right)^2\frac{(\gamma+1)(2\gamma-1)}{(2\gamma+1)}\right], \quad (3.26)$$

where the result is written for the dimensionless quantity $\widetilde{\mu} = m_e \overline{C}_2$.

Upon substituting (3.25) into (3.9), taking the limit $r \to \infty$ and comparing the result with (1.2) we get the magnetic moment of the electron in the state $2s_{1/2}$:

$$\mu = \frac{e\hbar}{m_e c}\widetilde{\mu} \approx \frac{e\hbar}{2m_e c}\left(1-\frac{(Z\alpha)^2}{12}\right). \quad (3.27)$$

This magnetic moment differs from that of the electron in the state $1s_{1/2}$, the latter being given by (3.21).

We proceed now to the state $2p_{1/2}$ for which $n = 2$, $l = 1$, $j = m = \frac{1}{2}$, $n_r = 1$, $\kappa = 1$. In this state there exists orbital rotation which produces a magnetic field in addition to the field due to the electron magnetic moment. The functions $f(r)$ and $g(r)$ for this state calculated from Eq. (36.11) of [2] are

$$f(r) = \sqrt{m_e + \varepsilon}\, A r^{\gamma-1} e^{-\lambda r}\left[\frac{Z\alpha m_e}{\lambda}-2-\left(\frac{Z\alpha m_e}{\lambda}-1\right)\frac{2\lambda r}{2\gamma+1}\right],$$

$$g(r) = -\sqrt{m_e - \varepsilon}\, A r^{\gamma-1} e^{-\lambda r}\left[\frac{Z\alpha m_e}{\lambda}-\left(\frac{Z\alpha m_e}{\lambda}-1\right)\frac{2\lambda r}{2\gamma+1}\right] \quad (3.28)$$

with

$$A^2 = \frac{(2\gamma+1)(2\lambda)^{2\gamma+2}}{8Z\alpha m_e^2\left(\dfrac{Z\alpha m_e}{\lambda}-1\right)\Gamma(2\gamma+1)}. \quad (3.29)$$

It follows from (2.8) – (2.9) for the present state that $Q = 0$ and

$$j_\varphi = P = i e^{-i\varphi} fg\sqrt{\frac{8}{3}}\, Y_{11}Y_{00} = \frac{1}{2\pi}fg\sin\vartheta. \quad (3.30)$$

This $j_\varphi$ differs from (3.7) in the sign. Therefore, instead of (3.12) we now have

$$\frac{dC_1}{dr} = -\frac{2}{3}fg, \quad \frac{dC_2}{dr} = \frac{2}{3}r^3 fg. \quad (3.31)$$

Integrating the first equation of (3.31) upon making use of (3.28) and (3.2) yields

$$C_1 = \frac{2\lambda}{3}A^2\left[\frac{Z\alpha m_e}{\lambda}\left(\frac{Z\alpha m_e}{\lambda}-2\right)I_{2\gamma-2}-\left(\frac{Z\alpha m_e}{\lambda}-1\right)^2\frac{4\lambda}{2\gamma+1}I_{2\gamma-1}+\left(\frac{Z\alpha m_e}{\lambda}-1\right)^2\frac{4\lambda^2}{(2\gamma+1)^2}I_{2\gamma}\right]$$

$$(3.32)$$

with the constant of integration $\overline{C}_1 = 0$. Upon integrating the second equation of (3.31) we obtain according to (3.11)

$$\zeta(r) = C_1(r)r$$

$$-\frac{2\lambda A^2}{3r^2}\left[\frac{Z\alpha m_e}{\lambda}\left(\frac{Z\alpha m_e}{\lambda} - 2\right)I_{2\gamma+1} - \left(\frac{Z\alpha m_e}{\lambda} - 1\right)^2\frac{4\lambda}{2\gamma+1}I_{2\gamma+2} + \left(\frac{Z\alpha m_e}{\lambda} - 1\right)^2\frac{4\lambda^2}{(2\gamma+1)^2}I_{2\gamma+3}\right] + \frac{\overline{C}_2}{r^2}$$

(3.33)

As before the constant $\overline{C}_2$ is to be found from the requirement that $\zeta(r)$ be finite as $r \to 0$. In view of (3.16) and (3.29) this entails

$$\widetilde{\mu} = \frac{\lambda(2\gamma+1)}{6Z\alpha m_e\left(\dfrac{Z\alpha m_e}{\lambda} - 1\right)}\left[\left(\frac{Z\alpha m_e}{\lambda} - 1\right)^2\frac{(\gamma+1)(2\gamma-1)}{2\gamma+1} - \frac{Z\alpha m_e}{2\lambda}\left(\frac{Z\alpha m_e}{\lambda} - 2\right)(2\gamma+1)\right], \quad (3.34)$$

where the result is written for the dimensionless quantity $\widetilde{\mu} = m_e\overline{C}_2$.

Comparing the last formulae with (1.2) as above we find the magnetic moment in the state $2p_{1/2}$

$$\mu = \frac{e\hbar}{m_e c}\widetilde{\mu} \approx \frac{e\hbar}{6m_e c}\left(1 - \frac{(Z\alpha)^2}{4}\right). \quad (3.35)$$

According to the outset of this section the orbital angular momentum operator $\mathbf{L} = [\mathbf{r}\cdot\mathbf{p}]$ has the same form as in the classical case. Therefore the magnetic moment due to the orbital rotation alone can be computed with the help of Eq. (44.5) of [1]. In our case

$$\mu = \frac{e\hbar m}{2m_e c}, \quad (3.36)$$

where $m = 0, \pm 1$ once $l = 1$. It is worthy of remark that, if $m = 0$, the wave function does not depend on the angle $\varphi$ which amounts to saying that the projection of the angular momentum on the z-axis in this case is zero. Comparing Eqs. (3.27), (3.35) and (3.36) we can see that there is no simple relation between the electron magnetic moment and the orbital magnetic moment taken separately in a state where the orbital rotation is present. We shall return to this question at the end of the section.

We consider next the state $2p_{3/2}$ for which $n = 2$, $l = 1$, $j = 3/2$, $n_r = 0$, $\kappa = -2$, taking $m = 3/2$. Since $n_r = 0$, we have (3.3) and (3.6) for $f(r)$ and $g(r)$. It follows from (2.8) – (2.9) that $Q = 0$ here and

$$j_\varphi = P = ie^{-3i\varphi}fg\frac{4}{\sqrt{5}}Y_{11}Y_{22} = -\frac{3}{4\pi}fg\sin^3\vartheta. \quad (3.37)$$

Equation (2.11) now yields

$$\frac{1}{r^2}\frac{\partial}{\partial r}\left(r^2\frac{\partial A_\varphi}{\partial r}\right)+\frac{1}{r^2}\frac{\partial}{\partial\vartheta}\left[\frac{1}{\sin\vartheta}\frac{\partial}{\partial\vartheta}\left(\sin\vartheta A_\varphi\right)\right]=3efg\sin^3\vartheta\,. \tag{3.38}$$

We look for a solution to the equation in the form

$$A_\varphi=e\zeta(r)\sin\vartheta+e\zeta_1(r)\sin^3\vartheta\,, \tag{3.39}$$

and obtain the following equations

$$\frac{d}{dr}\left(r^2\frac{d\zeta_1}{dr}\right)-12\zeta_1=3r^2fg\,,\quad \frac{d}{dr}\left(r^2\frac{d\zeta}{dr}\right)-2\zeta=-8\zeta_1\,. \tag{3.40}$$

The solution of the first of these equations with use made of the method of variation of constants can be represented as

$$\zeta_1(r)=C_1(r)r^3+\frac{C_2(r)}{r^4} \tag{3.41}$$

with

$$\frac{dC_1}{dr}=\frac{3}{7r^2}fg\,,\quad \frac{dC_2}{dr}=-\frac{3}{7}r^5fg\,. \tag{3.42}$$

Integrating the first equation with (3.3) for $f(r)$ and $g(r)$ yields

$$C_1=-\frac{3}{7}\lambda A^2 I_{2\gamma-4}\,, \tag{3.43}$$

where the constant of integration is put equal to zero; otherwise $\zeta_1(r)$ of (3.41) will increase with $r$.

Integration of the second equation of (3.42) produces

$$C_2=\frac{3}{7}\lambda A^2 I_{2\gamma+3}+\overline{C}_2\,. \tag{3.44}$$

The function $\zeta_1(r)$ of (3.41) becomes now

$$\zeta_1(r)=-\frac{3}{7}\lambda A^2 r^3 I_{2\gamma-4}+\frac{3}{7}\lambda A^2\frac{I_{2\gamma+3}}{r^4}+\frac{\overline{C}_2}{r^4}\,. \tag{3.45}$$

The constant $\overline{C}_2$ found from the requirement that $\zeta_1(r)$ be finite as $r\to 0$ is

$$\overline{C}_2=\frac{3(2\gamma+1)(2\gamma+2)(2\gamma+3)}{112\lambda^2}\approx\frac{45}{2(Z\alpha)^2}\left(1-\frac{107}{420}(Z\alpha)^2\right)\,. \tag{3.46}$$

Here Eq. (3.2) with $\kappa^2=4$ is taken into account.

The behaviour of $A_\varphi$ of (3.39) proportional to $1/r^4$ at infinity in view of (3.45) is characteristic of an octupole magnetic moment. To elucidate whether the octupole moment can be attributed to orbital rotation we calculate the current density relevant to the orbital rotation with use made of formulae of [3] for $n=2$, $l=m=1$:

$$j_\varphi = \frac{e\hbar}{64\pi m_e} \widetilde{r} e^{-\widetilde{r}} \sin\vartheta \,, \qquad (3.47)$$

where $\widetilde{r}$ is a dimensionless quantity expressed in the atomic units. The structure of this $j_\varphi$ is similar to the structure of $j_\varphi$ of (3.7) and leads to the dipole magnetic moment of (3.36), and nothing more. Thus the octupole magnetic moment in the state under consideration is completely due to the magnetic field of the electron alone.

Having considered the first equation of (3.40) we are coming now to the second equation whose solution is of the form

$$\zeta(r) = C_3(r)r + \frac{C_4(r)}{r^2} \qquad (3.48)$$

with

$$\frac{dC_3}{dr} = -\frac{8\zeta_1(r)}{3r^2}, \quad \frac{dC_4}{dr} = \frac{8r\zeta_1(r)}{3} \,. \qquad (3.49)$$

When integrating these equations one meets with integrals

$$I_p^{(n)}(r) = \int_\infty^r x^n I_p(x)dx = -\frac{1}{n+1}I_{p+n+1}(r) + \frac{r^{n+1}}{n+1}I_p(r) \,. \qquad (3.50)$$

With this formula at hand we integrate the equations of (3.49) and insert the result into (3.48):

$$\zeta(r) = \frac{12}{35}\lambda A^2 r^3 I_{2\gamma-4} - \frac{4}{5}\lambda A^2 r I_{2\gamma-2} + \frac{4}{5}\lambda A^2 r \frac{I_{2\gamma+1}}{r^2} - \frac{12}{35}\lambda A^2 r \frac{I_{2\gamma+3}}{r^4} - \frac{4\overline{C}_2}{5r^4} + \frac{\overline{C}_4}{r^2} \,. \qquad (3.51)$$

The terms in $1/r^4$ vanish as $r \to 0$ owing to (3.46). In the same limit the terms in $1/r^2$ give

$$\overline{C}_4 = \frac{2\gamma+1}{5m_e} \,. \qquad (3.52)$$

Proceeding as before we find the magnetic moment in the state $2p_{3/2}$ with $m = 3/2$:

$$\mu = \frac{e\hbar}{5m_e c}(2\gamma+1) \approx \frac{e\hbar}{m_e c}\left(1 - \frac{(Z\alpha)^2}{10}\right). \qquad (3.53)$$

To get a full picture for $n = 2$ we adduce without derivation the magnetic moment in the state $2p_{3/2}$ with $m = 1/2$:

$$\mu = \frac{e\hbar}{15m_e c}(2\gamma+1) \approx \frac{e\hbar}{3m_e c}\left(1 - \frac{(Z\alpha)^2}{10}\right). \qquad (3.54)$$

In this state there is an octupole magnetic moment too as above.

The results as given in Eqs. (3.35), (3.53) and (3.54), and compared with (3.36) demonstrate that, when the orbital rotation is present, the magnetic moment of the atom is modified in a great extent. This is possible only if the electron is not a tiny object but it occupies a substantial region in the atom and its structure changes depending on its state.

# 4. The vacuum polarization

The vacuum polarization modifies the electromagnetic potential. In this connection the question arises as to whether the vacuum polarization can alter the asymptotic behaviour of the vector potential given in (1.2). The electromagnetic potential $A_\mu$ with account taken of the vacuum polarization is written out in Eq. (14.109) of [4]:

$$A_\mu(\mathbf{x}) = -\int \frac{j_\mu^{\mathrm{ren,b}}(\mathbf{x}')}{4\pi \mid \mathbf{x} - \mathbf{x}' \mid}\left(1 + \frac{\alpha}{3\pi} Z_0 \mid \mathbf{x} - \mathbf{x}' \mid\right)d^3x', \qquad (4.1)$$

where $j_\mu^{\mathrm{ren,b}}(\mathbf{x}')$ is the renormalized current. Effects due to the vacuum polarization are reflected in the function $Z_0|\mathbf{x} - \mathbf{x}'|$ which falls off exponentially as $|\mathbf{x} - \mathbf{x}'| \to \infty$. An analogical result for the Coulomb potential follows from Eq. (114.7) of [2]. There are calculations [5] demonstrating that a function of the type $Z_0|\mathbf{x} - \mathbf{x}'|$ falls off proportionally to $1/|\mathbf{x} - \mathbf{x}'|^4$ as $|\mathbf{x} - \mathbf{x}'| \to \infty$.

In any case the asymptotic behaviour of the vector potential $\mathbf{A}$ embodied in Eq. (1.2) remains unchanged if the vacuum polarization is taken into account. It should be underlined that, although the vacuum polarization does not affect the asymptotic behaviour of $\mathbf{A}$, it can alter the magnetic moment $\mu$ itself. However, a contribution to $\mu$ owing to the vacuum polarization is proportional at least to $\alpha$ as in (1.1) and thereby the contribution cannot essentially modify the magnetic moments of (3.35), (3.53) and (3.54), and the octupole magnetic moment following from (3.46).

# 5. Concluding remarks

There exist laws of conservation of charge and of angular momentum. As a result, no external actions can change the charge and spin of the electron. If, for example, one calculates the electrostatic potential $\varphi 2(r)$ created by the electron in all the states considered in Sec. 3, one will inevitably obtain that $\varphi(r) \to e/r$ as $r \to \infty$. At the same time there is no law of conservation of magnetic moment. Consequently, if the electron is not a point-like particle and is deformable (see Introduction), its magnetic moment may be different in different situations. The results of the present paper demonstrating that the magnetic moment of the electron is dissimilar in different fields and states indicate that the electron is not a point-like particle and is deformable. Moreover, the electron is not a tiny particle but it occupies a substantial region in the atom.

Some comment at this point is in order concerning the customary probabilistic interpretation of the wave function according to which $|\psi|^2 dV$ is the probability of finding a particle in an infinitesimal volume $dV$. It should be emphasized that the particle in this interpretation must be regarded as point-like otherwise one cannot speak about the particle in the infinitesimal volume $dV$. If the electron were point-like, the asymptotic form of the vector potential created by the electron would remain the same, be the electron immobile or be it moving in an atom. This can be shown as follows. Let $\mathbf{a}$ of magnitude $a$ be the vector that determines the position of the electron with respect to the origin $\mathbf{r} = 0$. Let $\mathbf{r}'$ of magnitude $r'$ be the vector from the electron to a point of observation with the radius vector $\mathbf{r}$. By analogy with (1.2) the asymptotic form of the vector potential created by the electron will be

$$A_\varphi = \frac{\mu}{r'^2} \sin\vartheta', \tag{5.1}$$

in which $\vartheta'$ is the angle between the z-axis and the vector $\mathbf{r}'$. Seeing that $\mathbf{r} = \mathbf{a} + \mathbf{r}'$, we have

$$r'^2 = r^2 - 2ra\cos\beta + a^2, \tag{5.2}$$

where $\beta$ is the angle between $\mathbf{r}$ and $\mathbf{a}$. It can be demonstrated that

$$\sin\vartheta' = \sin\vartheta\left(1 - \frac{a^2}{2r^2}\sin^2\beta\right) \pm \frac{a}{r}\sin\beta\left(1 + \frac{a}{2r}\cos\beta\right)\cos\vartheta + O\left(\frac{a^3}{r^3}\right). \tag{5.3}$$

We substitute these into (5.1). In experiment one observes averaged quantities. In the absence of orbital rotation the average value of $\sin\beta$ and $\cos\beta$ is zero whereas the average value of $\sin^2\beta$ and $\cos^2\beta$ is equal to ½. We average (5.1) over $\beta$ and $a$ with the result

$$A_\varphi = \frac{\mu}{r^2}\sin\vartheta\left[1 + \frac{15\overline{a^2}}{4r^2} + O\left(\frac{a^3}{r^3}\right)\right], \tag{5.4}$$

where $\overline{a^2}$ is the average value of $a^2$. The magnitude of $a$ is of the order of the Bohr radius $a_{\mathrm{B}} = \hbar^2/(me^2Z)$. The noteworthy fact is that $a_{\mathrm{B}}$ is independent of $\alpha = e^2/\hbar c$ because $a_{\mathrm{B}}$ does not contain $c$. It follows from the asymptotic form of (5.4) that the electron magnetic moment $\mu$ should be the same in all states where the orbital motion is absent (the above states $1s_{1/2}$ and $2s_{1/2}$) and should be equal to $\mu$ of an immobile electron where $a = 0$. Moreover, in the states $1s_{1/2}$ and $2s_{1/2}$ there must be an octupole magnetic moment owing to the term with $\overline{a^2}$ in (5.4). All of these are in sharp contrast to the calculations in Sec. 3. This reasoning too confirms the fact that the electron cannot be point-like.

In Ref. [6] a new formulation of QED was proposed in which the electronic and electromagnetic fields are ordinary $c$-numbers in contradistinction to noncommuting $q$-numbers used in the standard formulation of QED. Instead of the probabilistic interpretation of the wave function the quantity $|\psi|^2$ of [6] represents a real distribution of the electronic density. According to the latter interpretation the electron looks like a cloud filling the region where $|\psi|^2 \neq 0$. From this point of view it is easy to understand the results of Sec. 3. The Coulomb field of the nucleus of an atom changes the shape of the electronic cloud, which gives rise to modification of its magnetic moment. The deformation of the electronic cloud is particularly pronounced in states where the quantum number $l \neq 0$, resulting in a significant modification of the magnetic moment according to Eqs. (3.35), (3.53), (3.54) and (3.46).

It would be of interest to calculate the magnetic moment of a free electron. This can be done with the help of equations of [6]. However those equations constitute a set of six intricate nonlinear differential equations which can be solved only numerically. But numerical solution of the set presents a challenging mathematical problem.

A word should be added about the electron considered to be a point-like particle. It is believed thus far that the electron size is so small that various attempts to find it have not met with success. These attempts prove, however, that the electron has no foreign core where unknown forces act and nothing more. It should be emphasized that nowhere in QED does one regard the electron as point-like. The point-like particle is represented by a delta function. Nowhere in QED does one represent the electron by the delta function. For example, when one considers scattering of a photon by the electron, one represents the electron by a density matrix which is not the delta function, see § 86 of [2]. The electron in Feynman diagrams is not represented by the delta function either. Could the electron be described with the help of the delta function, this would greatly simplify involved calculations in QED.

According to [6] the free electron has the size lying in the range between the electron Compton wavelength and the Bohr radius. It is explained in [6] that not only does such a large size of the electron not contradict experiment, but on the contrary it is supported by all experimental data concerning quantum mechanics.